\documentclass[a4paper, 12pt, onecolumn]{article}

\usepackage[cmex10]{amsmath}	% cmex10: nutzt nur T1 fonts, vermeidet Bitmaps! (bspw. in Fußnoten)
\usepackage{amssymb,amsthm,mathtools}
\mathtoolsset{showonlyrefs}

\usepackage[latin1]{inputenc}
\usepackage[english]{babel}

\usepackage[T1]{fontenc}
\usepackage{times}

\usepackage{booktabs}
\usepackage{enumitem}

\usepackage[labelfont=bf]{caption}

\usepackage[ruled,linesnumbered]{algorithm2e}
\DontPrintSemicolon
\SetFuncSty{texttt}
\SetArgSty{ensuremath}
\SetProcNameSty{texttt}
\SetProcArgSty{ensuremath}
\SetKwInOut{Input}{Input}
\SetKwInOut{Output}{Output}
\SetKwInOut{Dummy}{Out\hspace{1pt}put} % verhindert einen Bug, bei dem der Doppelpunkt nach dem längsten, mit SetKwInOut definiertem Schlüsselwort in einer eigenen Zeile erscheint.

\usepackage[nolist]{acronym}

\begin{acronym}
	\acro{PPM}{Prediction by Partial Matching}
	\acro{DMC}{Dynamic Markov Coding}
	\acro{CTW}{Context Tree Weighting}
	\acro{PAQ}{``Pack''}
	\acro{AC}{Arithmetic Coding}
	\acro{CM}{Context Mixing}
	\acro{CG}{Conjugate Gradient}
	\acro{KT}{Krichevsky-Trofimov}
	\acro{iid}{independent identically distributed}
	\acro{BWT}{Burrows-Wheeler-Transform}
	\acro{BFGS}{Broyden-Fletcher-Goldfab-Shanno}
	\acro{KKT}{Karush-Kuhn-Tucker}
	\acro{WFC}{Weighted Frequency Counting}
	\acro{MTF}{Move-to-Front}
	\acro{LP}{Laplace}
	\acro{SAKDC}{Swiss Army Knife Data Compression}
	\acro{SQP}{Sequential Quadratic Programming}
	\acro{bpc}{bits per character}
  \acro{OGD}{Online Gradient Descent}
  \acro{BW}{Beta-Weighting}
  \acro{pd}{probability distribution}
  \acro{PTW}{Partition Tree Weighting}
\end{acronym}

\ifdefined\Compact{}
  \usepackage[margins=normal,bibliography=normal,lists=normal]{savetrees} % oder: 	all=normal,title=tight,sections=tight
  \usepackage{thmtools}
  \setenumerate[1]{noitemsep,topsep=2pt}

  \renewenvironment{thebibliography}[1]{%
    \begin{oldthebibliography}{#1}%
      \small
      \setlength\parsep{0pt}%
      \setlength\itemsep{-.9ex}%
  }{%
     \end{oldthebibliography}%
  }

  \newenvironment{prf}[1][Proof]
  {\vspace{-0.75\topsep}\begin{proof}[#1]}{\end{proof}\vspace{-.75\topsep}}
  
  \newtheoremstyle{plain1}
    {0.3\topsep}   % ABOVESPACE, 0.5\topsep für remark
    {0.3\topsep}   % BELOWSPACE, 0.5\topsep für remark
    {\itshape}  % BODYFONT, \normalfont für remark, def.
    {}       % INDENT (empty value is the same as 0pt)
    {\bfseries} % HEADFONT, \itshape für remark
    {.}         % HEADPUNCT
    {5pt plus 1pt minus 1pt} % HEADSPACE
    {}          % CUSTOM-HEAD-SPEC
  \newtheoremstyle{definition1}
    {0.3\topsep}   % ABOVESPACE
    {0.3\topsep}   % BELOWSPACE
    {\normalfont}  % BODYFONT
    {}       % INDENT (empty value is the same as 0pt)
    {\bfseries} % HEADFONT
    {.}         % HEADPUNCT
    {5pt plus 1pt minus 1pt} % HEADSPACE
    {}          % CUSTOM-HEAD-SPEC
  \newtheoremstyle{remark1}
    {0.3\topsep}   % ABOVESPACE, 0.5\topsep für remark
    {0.3\topsep}   % BELOWSPACE, 0.5\topsep für remark
    {\normalfont}  % BODYFONT, \normalfont für remark, def.
    {}       % INDENT (empty value is the same as 0pt)
    {\itshape} % HEADFONT, \itshape für remark
    {.}         % HEADPUNCT
    {5pt plus 1pt minus 1pt} % HEADSPACE
    {}          % CUSTOM-HEAD-SPEC

  \newcommand{\plaintheoremstyle}{\theoremstyle{plain1}}
  \newcommand{\remarkstyle}{\theoremstyle{remark1}}
  \newcommand{\definitionstyle}{\theoremstyle{definition1}}
    
  \AtBeginDocument{
    \setlength\abovedisplayskip{3pt plus 2pt}
    \setlength\belowdisplayskip{3pt plus 2pt}
    \setlength\abovedisplayshortskip{3pt plus 2pt}
    \setlength\belowdisplayshortskip{3pt plus 2pt} 
    
    \setlength{\floatsep}{6pt}
    \setlength{\textfloatsep}{6pt}
    \setlength{\intextsep}{6pt}
  }
\else 

  \newenvironment{prf}[1][Proof]{\begin{proof}[#1]}{\end{proof}}
  \newcommand{\plaintheoremstyle}{\theoremstyle{plain}}
  \newcommand{\remarkstyle}{\theoremstyle{remark}}
  \newcommand{\definitionstyle}{\theoremstyle{definition}}
\fi

\usepackage[top=1in, left=1.25in, textwidth=6in, textheight=9in]{geometry} 	
\plaintheoremstyle
\newtheorem{thm}{Theorem}[section]
\newtheorem{prp}[thm]{Proposition}
\newtheorem{lem}[thm]{Lemma}

\remarkstyle

\definitionstyle
\newtheorem{dfn}[thm]{Definition}
\newtheorem{ex}[thm]{Example}
\newtheorem{obs}[thm]{Observation}

\renewcommand{\P}{{\mathcal{P}}}	% überschreibt Makro 
\renewcommand{\S}{{\mathcal{S}}}	% überschreibt Makro für "§"
\newcommand{\I}{{\mathcal{I}}}
\newcommand{\R}{{\mathcal{R}}}

\newcommand{\X}{{\mathcal{X}}}

\newcommand{\eps}{{\varepsilon}}

\newcommand{\rfd}{\textsc{rfd}}
\newcommand{\mdl}{\textsc{mdl}}

\newcommand{\block}[1]{\smallskip\par\noindent\makebox{\textsl{#1}}}

\title{\large{On Probability Estimation via Relative Frequencies and Discount}\ifdefined\Compact{}\,\footnote{The full version of this paper is available at \texttt{http://arxiv.org/abs/1311.1723}.}\fi}
\author{
	\vspace{2mm}
	Christopher Mattern \\ 
	Technische Universität Ilmenau\\
	Ilmenau, Germany \\
	\texttt{christopher.mattern@tu-ilmenau.de}
}
\date{}	% kein Datum anzeigen

\begin{document}

\maketitle
\thispagestyle{empty}	% Seitennummer auf 1. Seite unterdrücken (troz pagestyle empty!)

%%%%%%%%%%%%%%%%%%%%%%%%%%%%%%%%%%%%%%%%%%%%%%%%%%%%%%%%%%%%%%%%%%%%%%%%%%%%%%%%

%%% Abstrakt %%%

\ifdefined\Compact{}
\vspace*{-30pt}
\fi
% \begin{abstract}
% \noindent
% Probability estimation is an elementary building block of every statistical data compression algorithm.
% In practice probability estimation is often based on relative letter frequencies which get scaled down, when their sum is too large.
% Such algorithms are attractive in terms of memory requirements, running time and practical performance.
% However, there still is a lack of theoretical understanding.
% In this work we formulate a typical probability estimation algorithm based on relative frequencies and frequency discount, Algorithm~$\rfd$.
% Our main contribution is its theoretical analysis.
% We show that the code length it requires above an arbitrary piecewise stationary model with bounded and unbounded letter probabilities is small.
% This theoretically confirms the recency effect of periodic frequency discount, which has often been observed empirically.
% \end{abstract}
\begin{center}
\parbox{.9\textwidth}{
\small
\paragraph{Abstract.}
Probability estimation is an elementary building block of every statistical data compression algorithm.
In practice probability estimation is often based on relative letter frequencies which get scaled down, when their sum is too large.
Such algorithms are attractive in terms of memory requirements, running time and practical performance.
However, there still is a lack of theoretical understanding.
In this work we formulate a typical probability estimation algorithm based on relative frequencies and frequency discount, Algorithm~$\rfd$.
Our main contribution is its theoretical analysis.
We show that Algorithm~$\rfd$ performs almost as good as any piecewise stationary model with either bounded or unbounded letter probabilities.
This theoretically confirms the recency effect of periodic frequency discount, which has often been observed empirically.
}
\end{center}

%%%%%%%%%%%%%%%%%%%%%%%%%%%%%%%%%%%%%%%%%%%%%%%%%%%%%%%%%%%%%%%%%%%%%%%%%%%%%%%%

\section{Introduction}

\paragraph{Background.}
Sequential probability assignment is an elementary component of every statistical data compression algorithm, such as \ac{PPM}, \ac{CTW}, \ac{PAQ} and \ac{DMC}.
Statistical compression algorithms split compression into modeling and coding and process an input sequence symbol-by-symbol.
During modeling a model computes a distribution $p$ and during coding an encoder maps the next character $x$, given $p$, to a codeword of a length close to $-\log p(x)$ bits (this is the ideal code length).
Decoding is the reverse: Given $p$ and the codeword the decoder restores $x$.
Functions \FuncSty{Compress} and \FuncSty{Decompress} in Figure~\ref{alg:rfd} illustrate this process.
\ac{AC} is the de facto standard en-/decoder, it closely approximates the ideal code length \cite{eoit}.
All of the mentioned compression algorithms require simple, elementary models to predict a probability distribution.
Elementary models are typically based on relative letter frequencies, which get discounted, when their sum is too large.
Out of a huge amount of elementary models only a small subset, selected by a context, contributes to the prediction of a statistical compression algorithm in a step.
Nevertheless, elementary models have a big impact on both theoretical guarantees \cite{cts12, ctw_95} and empirical performance \cite{cm_ccp2011,ptw13} of statistical compression algorithms.
Typically, we express theoretical guarantees on a model by the amount of bits the model requires above an ideal competing scheme (e.\,g. a fixed probability distribution) assuming ideal encoding.
This code length excess is called \emph{redundancy} w.\,r.\,t. the competing scheme.

\paragraph{Previous Work.}
Relative frequency-based elementary models, such as the Laplace- and KT-Estimator, are well-known and well-understood \cite{eoit}. 
To our knowledge refinements thereof that periodically discount letter frequencies were first analyzed in \cite{howard91} by Howard and Vitter.
They assume that the frequency for letter $x$ is increased by $1$ after we observe $x$.
After a block of $B$ letters, frequencies get multiplied by a constant $0<f<1$ (rescaling).
In the analysis of this scheme the code length is characterized in terms of weighted entropy.
Unfortunately the weighted entropy is not directly related to a competing scheme, rather it is based on the algorithm itself.
This makes it hard to interpret the results.
Furthermore, frequency increments other than $1$ (even variable increments, selected by a small context) can result in an improved empirical performance \cite{ppm_ii}.
Another drawback in the setting of \cite{howard91} is that rescales happen every $B$ letters.
Thus an additional counter for every elementary model is required.
Since a statistical compressor typically uses millions of elementary models, a compact design of an elementary model is desirable \cite{ppm_ii}.
Clearly, discounting frequencies gives more weight to recent statistics.
So discarding should intuitively reduce the redundancy w.\,r.\,t. an adaptive competitor, such as a sequence of fixed probability distributions.
For brevity, we will term such a competitor piecewise stationary model, \textsc{pws}.
In this paragraph let $s$ denote the number of probability distributions of an arbitrary, fixed \textsc{pws} and let $n$ be the length of the sequence to be compressed.
To achieve low redundancy w.\,r.\,t. \textsc{pws} several authors proposed mixtures over elementary models associated to a so-called transition diagram \cite{shamir99,lad,willems96}.
The idea was introduced by Willems \cite{willems96}.
This approach is the converse of the simpler scheme of frequency discount.
The most advanced algorithms from this family are due to Shamir \cite{shamir99}.
Shamir proposed three algorithms with total running time ranging from $O(n)$ to $O(n^2)$ and redundancy w.\,r.\,t. \textsc{pws} ranging from $O(s \log n)$ to $O(s \sqrt{n\log n} )$.
The simplest algorithm of \cite{shamir99} collapses to a KT-estimator which is reset periodically and has redundancy $O(s \sqrt{n\log n})$.
The intervals between resets have increasing length.
In practice a reset of statistics is undesirable, it typically hurts compression compared to a rescale (which retains a certain amount of old statistics).
Recently, Veness proposed \ac{PTW}, an algorithm framework to deal with \textsc{pws} \cite{ptw13}.
\ac{PTW} inherits its guarantees from the \ac{CTW} method.
If we employ the KT-estimator in this framework, it achieves redundancy $O(s\log^2 n)$ and has running time $O(n\log n)$.
Unfortunately, most of the mentioned algorithms share time (and space) requirements beyond practical scope, namely total running time $O(n)$.

\paragraph{Our Contribution.}
In this work we present a practical, widely used and intuitive elementary model and bound its redundancy w.\,r.\,t. \textsc{pws} with bounded and unbounded probability on any letter.
Our work is much in spirit of \cite{howard91}, but provides results that are easier to interpret, similar to \cite{shamir99}.
We provide the first analysis that theoretically confirms the intuitive relation between frequency discount (not just a complete reset, as in \cite{shamir99}) and adaptivity, in other words low redundancy w.\,r.\,t. \textsc{pws}.
Our elementary model requires $O(1)$ operations per letter and $O(N)$ frequency counters, for an alphabet of size $N$.
By choosing parameters of our algorithm appropriately (see Example~\ref{ex:L}) we obtain redundancy $O(s \sqrt n \log n)$ w.\,r.\,t. \textsc{pws} with $s$ probability distributions.
The practitioner's approach to modeling is above a similar theoretician's approach of \cite{shamir99} only by a factor $O(\sqrt{\log n})$.

We divide the remaining part of the work as follows.
Section~\ref{sec:preliminaries} introduces our notation, the algorithm of interest, its parameters and conditions on the parameters, which we assume for our analysis.
Next, Section~\ref{sec:closeups} investigates at which intervals rescales takes place.
(This turns out be be important for later analysis.)
In Section~\ref{sec:analysis} we provide redundancy bounds and subsequently refine these to yield our two main results.
Furthermore, we interpret the influence of the parameters on the redundancy bounds.
Finally, Section~\ref{sec:conclusion} summarizes our work and gives possible topics for future research.

\section{Preliminaries}\label{sec:preliminaries}

\paragraph{Notation.}
Calligraphic letters denote sets, typewriter style declares variables and function names.
Unless stated differently, intervals such as $(a,b]$ and $[a,b]$ are intervals of integers.
A partition of an interval $(a,b]$ of integers is a set of $s\geq1$ disjoint intervals $(i_0, i_1], \dots, (i_{s-1}, i_s]$ of integers, such that $i_0:=a<i_1<\dots<i_s:=b$.
For $1\leq k\leq s$ the term ``$k$-th segment'' uniquely referes to $(i_{k-1}, i_k]$.
For integers $j\geq i$ an array with cells $i$, $i+1$, \dots, $j$ is $\texttt A[i..j]$ and $\texttt A[k]$ accesses array cell $k$.
Let $\X:=\{1, 2,\dots, N\}$ be an alphabet of cardinality $1<N<\infty$ and let $x_a^b := x_a x_{a+1} \dots x_b$ be a sequence over $\X$ where $x^n$ abbreviates $x_1^n$.
The set of all probability distributions over $\X$ is $\P$.
%The set of all probability distributions over $\X$ with non-zero probability on all letters is $\P_+$, with probability at least $\eps>0$ on all letters is $\P_\eps$ and $\P:=\P_0$ (we set $\eps=0$).
The natural logarithm is ``$\ln$'', whereas ``$\log$'' is the base-two logarithm.
For $x\in\X$ and $p\in\P$ we denote the (ideal) binary code length of $x$ w.r.t. $p$ as $\ell(x, p) := -\log p(x)$ and define $\ell(x^n; p) := \sum_{1\leq k\leq n} \ell(x_k; p)$.
For $p,q\in\P$, $H(p)$ is the entropy of $p$ and $D(p\parallel q)$ is the KL-distance between $p$ and $q$, as defined in \cite{eoit}.

\paragraph{Models.}
As mentioned earlier, the model of a statistical data compression algorithm is of central importance and greatly determines compression.
We now define it formally.

\begin{dfn}
A \emph{model} $\mdl$ maps a sequence $x^k$ of length $k\geq0$ over $\X$ to a probability distribution $p\in\P$. % with non-zero probability on any letter.
%model distribution $p\in\P_+$.
We write $\mdl(x; x^k)$ for $p(x)$.
Model $\mdl$ assigns $\ell(x^n; \mdl) := - \sum_{1\leq k\leq n} \log \mdl(x_k; x^{k-1})$ bits to sequence $x^n$.
Models must not give probability $0$ to a letter:
If $\mdl(x_k; x^{k-1})=0$ for any $1\leq k\leq n$, then $\ell(x^n; \mdl):=\infty$.
%\TODO{$p(x)>0$ fordern und zwischen "`model"' und "`competitor"' unterscheiden?}
\end{dfn}

\noindent
Typically statistical data compression algorithms combine and refine several ``elementary models''.
An ``elementary model'' is defined by a simple closed-form expression, such as the Laplace estimator $\mdl(x; x^n)=(\lvert\{1\leq k\leq n\mid x_k=x\}\rvert + 1)/(n+N)$.
(It is hard to precisely define what makes up an ``elementary model'', thus we stick with the informal notion just given.)

\begin{figure}[h!]
\begin{function}[H]  
  \caption{Compress() / \texttt{Decompress}}
  $\texttt{S}\gets \FuncSty{Init}()$\;
  \While{\text{not end{-}of{-}sequence}}{   
    $\texttt p \gets \FuncSty{Predict(\texttt S)}$\;    
    $\texttt x \gets $ read symbol (compression) \textbf{or} \FuncSty{Decode}(\texttt p) (decompression)\;
    \FuncSty{Encode}(\texttt p, \texttt x) (compression) \textbf{or} write symbol \texttt x (decompression)\;    
    $\texttt S \gets$\FuncSty{Update(\texttt S)}
  }
\end{function}
\begin{minipage}[!h]{0.5\textwidth}
\begin{function}[H]
  \caption{Update(\texttt s\lbrack 1..N\rbrack, \texttt t, \texttt x)}
  \If{\texttt t+d>T}{
    $\texttt t\gets0$\;\label{alg:rescale0}
    \For{\texttt i=1~\KwTo~N}{
      $\texttt s[\texttt i]\gets \lfloor c\cdot \texttt s[\texttt i]\rfloor$\;
      \lIf{\texttt s[\texttt i]=0}{
        $\texttt s[\texttt i]\gets 1$
      }
      $\texttt t \gets \texttt t + \texttt s[\texttt i]$\;
    }\label{alg:rescale1}
  }
  $\texttt s[\texttt x]\gets \texttt s[\texttt x]+d$\;\label{alg:update_s}
  $\texttt t \gets \texttt t +d$\; 
\end{function}
\end{minipage}
~
\begin{minipage}[!h]{0.48\textwidth}
\begin{function}[H]
  \caption{Predict(\texttt s\lbrack 1..N\rbrack, \texttt t)}
  create Array $\texttt p[1..N]$\;  
  \lFor{\texttt i=1~\KwTo~N}{$\texttt p[\texttt i]\gets \texttt s[\texttt i]/\texttt t$}
  \KwRet \texttt p\;  
\end{function}
\vspace{3.5pt}
\begin{function}[H]
  \caption{Init()}
  create Array $\texttt s[1..N]$\;  
  \lFor{i=1~\KwTo~N}{$\texttt s[i]\gets s_0(i)$}
  $t\gets s_0(1) + \dots + s_0(N)$\;  
  \KwRet (\texttt s, \texttt t)\;
\end{function}
\end{minipage}
\vspace{-12pt}
\renewcommand\figurename{\small Figure}
\caption{\small Compression, decompression and helper functions for statistical data compression.
Functions \FuncSty{Encode} and \FuncSty{Decode} are considered ideal.
Variable \texttt S in functions \FuncSty{Compress} and \FuncSty{Decompress} represents the state of the statistical model.
Functions \FuncSty{Init}, \FuncSty{Predict} and \FuncSty{Update} plugged into function \FuncSty{Compress} / \FuncSty{Decompress} make up Algorithm~$\rfd$.}
\label{alg:rfd}
\end{figure}
\vspace{-6pt}

\paragraph{Algorithm Relative Frequencies with Discount (\rfd).}
We now present an intuitive elementary model, \rfd, which is widely used in practice and will be analyzed in detail.
The following refers to the pseudocode given in Figure~\ref{alg:rfd}.
Algorithm~$\rfd$ is based on relative letter frequencies, which we store in an array $\texttt s[1..N]$ of integers.
Symbol $x$ has frequency $\texttt s[x]$.
For efficiency variable \texttt t stores the sum of all letter frequencies.
Thus, the state of $\rfd$ is uniquely determined by the pair $(\texttt s, \texttt t)$.
Naturally $\rfd$ assigns probability $\texttt s[x]/\texttt t$ to letter $x$ (see function \FuncSty{Predict}).
Since this quantity must be positive, $\rfd$ ensures that $\texttt s[x]\geq1$, at any time.
Now, after observing a new letter $x$ we must update the model (cf. function \FuncSty{Update}).
Therefore let $t$ be the content of variable \texttt t at the beginning of such an update.
If $t$ is not too large, in particular $t+d\leq T$, for some threshold $T$, we simply increase $\texttt s[x]$ and \texttt t by an integer $d$.
However, if $t$ is too large, we perform a so-called ``rescale'' before increasing $\texttt s[x]$ and \texttt t:
We replace the content of $\texttt s[i]$ by $\lfloor c\cdot \texttt s[i]\rfloor$ for some $0<c<1$ (e.\,g. an integer division), for all letters $i$.
As a fixup we assign value $1$ to any array cell which now has value $0$.
Variable \texttt t is now set to hold the sum of all array cells $\texttt s[i]$.
This completes the rescale and we can now increase $\texttt s[x]$ and \texttt t by $d$, as usual.
A rescale discounts the weight of old statistics and limits the number of bits for frequency storage to $O(N\log T)$ bits.
Clearly, this procedure has three parameters, namely $T$, $c$ and $d$.
Throughout this paper we assume the following on these parameters and on the initial values of \texttt s and \texttt t, given by function \FuncSty{Init}:
\begin{enumerate}[label={(C\arabic*)}]
  \item\label{it:cond_d}\label{it:cond0}
    Parameter $d\geq1$ is an integer.
    %Parameter $d$ is a multiple of the alphabet size $N$.
  \item\label{it:cond_c}
    Parameter $0\leq c<1$ is a real.
  \item\label{it:cond_T}
    Parameter $T$ is an integer and satisfies $d\leq (1-c)(T-N)$.
  \item\label{it:cond_s}\label{it:cond1}
    In \FuncSty{Init} constant $s_0(x)$ is a positive integer, for any letter $x$, and $\sum_{x\in\X}s_0(x)\leq T$.
\end{enumerate}

\section{Algorithm~$\rfd$ -- Closeups}\label{sec:closeups}

\paragraph{Mechanics of Algorithm \rfd.}
For a sequence $x^n$ the execution of function \FuncSty{Compress} (or \FuncSty{Decompress}) with the given functions \FuncSty{Init}, \FuncSty{Predict} and \FuncSty{Update} (see Figure~\ref{alg:rfd}) create a realization of Algorithm~$\rfd$, i.\,e. the sequence of operations:
\begin{align}
  (\texttt{\small s}, \texttt{\small t})\gets\FuncSty{\small Init}(),~
  &\texttt{\small p}\gets\FuncSty{\small Predict}(\texttt{\small s}, \texttt{\small t}),~
  (\texttt{\small s}, \texttt{\small t})\gets\FuncSty{\small Update}(\texttt{\small s}, \texttt{\small t}, x_1),~
  \dots,~\\  
  &\texttt{\small p}\gets\FuncSty{\small Predict}(\texttt{\small s}, \texttt{\small t}),~
  (\texttt{\small s}, \texttt{\small t})\gets\FuncSty{\small Update}(\texttt{\small s}, \texttt{\small t}, x_n)
  \text.
  \label{eq:opseq}
\end{align}
The phrase ``step $k$'' refers to the $k$-th pair of \FuncSty{Predict} and \FuncSty{Update} operations.
We say, a \emph{rescale} takes place in step $k$, if \FuncSty{Update} executes lines \ref{alg:rescale0} to \ref{alg:rescale1} in step $k$.
Rescales partition sequence \eqref{eq:opseq} into subsequences.
This makes up a \emph{rescale partition} $\R$ of $[1,n]$.
Since this view will become important in the analysis we introduce a formalization.

\begin{dfn}
A \emph{realization} of Algorithm~$\rfd$ consists of a sequence \eqref{eq:opseq} of operations with $n$ \FuncSty{Predict} (see Figure~\ref{alg:rfd}) operations and $n$ \FuncSty{Update} (see Figure~\ref{alg:rfd}) operations.
For $0\leq k< n$ we write $\rfd(x; x^k)$ for the content of array cell $\texttt p[x]$ in step $k+1$.
%(Array \texttt p is the return value of the $(k+1)$-th call to \FuncSty{Predict}.)
\end{dfn}

\begin{dfn}
For a realization of Algorithm~$\rfd$ and integers $1\leq a\leq b\leq n$ we call set $\R$ a \emph{rescale partition of $[a,b]$}, if 
$\R$ is a partition of $[a,b]$ and if for $[i,j]\in\R$ in steps $i, i+1, \dots, j$ there is exactly one rescale in step $j$ and $j<b$, or there is at most one rescale in step $j$ and $j=b$.
Whenever $a$ and $b$ are omitted, we have $a=1$ and $b=n$.
\end{dfn}

\noindent
Clearly, $\rfd(x; x^k)$ defines a model and can be interpreted as ``the prediction of Algorithm~$\rfd$ after processing input sequence $x^k$''.

\paragraph{Properties of a Rescale Partition.}
In the upcoming analysis of Algorithm $\rfd$ the structure of a rescale partition, in particular the length $m$ of a segment, will play an important role.
Thus, we will now take a closer look at it.
First, define
\begin{align}
  L := \frac{T-N}d
  \label{eq:L}
  \text,
\end{align}
which we will rely on shortly.
A rescale takes place whenever the content $t$ of variable \texttt t satisfies $t+d>T$.
Hence, if \texttt t has value $t$ after \FuncSty{Init} (or after the $k$-th rescale in \FuncSty{Update}), then the first (or $(k+1)$-th) rescale will approximately take place $m\approx(T-t)/d$ steps later.
So dealing with $m$ requires dealing with $t$, more formally:

\begin{lem}\label{lem:total_max}
For any realization of Algorithm $\rfd$, the content $t$ of variable \texttt{\upshape t} at the end of the $k$-th call to \FuncSty{Update}, for $k\geq0$ (for $k=0$ this is the situation after \FuncSty{Init}), satisfies
\[
  t\leq
  \begin{cases}    
    d+N+cdL, &\text{\upshape if a rescale takes place in step $k$}\\
    T,& \text{\upshape otherwise}    
  \end{cases}
  \text,
\]
where $d+N+cdL\leq T$.
\end{lem}
%\ifdefined\Compact{}
%\else
\begin{prf}
By Condition \ref{it:cond_T} we have $d+N+cdL\leq T$.
We prove the above claim by induction on $k$.
We now use induction on $k$ to prove the two bounds given above.
\block{Base. ---}
For $k=0$ we consider the situation right after \FuncSty{Init}.
No rescale takes place and we have $t=\sum_{x\in\X} s_0(x)\leq T$ by Condition \ref{it:cond_s}.
\block{Step. ---}
For $k>0$ let $t_0$ be the content of variable \texttt t and $s(x)$ be the content of $\texttt s[x]$, both at the beginning of step $k$.
We distinguish two cases:
\block{Case 1: No rescale takes place.}
We have $t_0+d\leq T$ and $t=t_0+d$.
\block{Case 2: A rescale takes place.}
We have $t_0+d>T$ and obtain
\[
  t = d + \sum_{x\in\X} \max\{1, \lfloor c\cdot s(x) \rfloor\} \leq d + N + c(t_0-N) \leq d + N + cdL
  \text.
\]
The first inequality is due to $\max\{1, \lfloor c\cdot s(x)\rfloor\}\leq c\cdot s(x) + 1-c$, for $s(x)\geq1$ and $0\leq c\leq1$ and
the last inequality follows from \eqref{eq:L} and $t_0\leq T$ (by the induction hypothesis).
% DCC version:
% Let $M$ be the number of symbols for which $c \cdot s(x)<2$.
% Then
% \begin{align}
  % t
  % &= d + \sum_{x\in\X} \max\{1, \lfloor c\cdot s(x) \rfloor\}
  % \leq d + M + \sum_{\substack{x\in\X\\ c\cdots(x)\geq 2 }} c\cdot s(x) \\
  % &= d + M + c t_0 - \sum_{\substack{x\in\X\\ c\cdots(x)<2 }} c\cdot s(x)
  % \leq d + M + c (T-M)\label{eq:total_max1}
  % \text.
% \end{align}
% The last step follows due to $t_0\leq T$ (by the induction hypothesis) and since $\sum_{c s(x)<2} s(x)\geq M$ (the content of array cell \texttt s[x] is at least $1$, at any time).
% By $M\leq N$ and \eqref{eq:L} the claim follows.
\end{prf}
%\fi

\noindent
By the upper bound on $t$ immediately after a rescale we can now bound $m$ (i.\,e. the length of a segment from $\R$, which is neither the first segment, nor the last segment) from below.

\begin{lem}\label{lem:segment_length}
For any realization of Algorithm $\rfd$ the segments of rescale partition $\R$ satisfy:
\begin{enumerate}[leftmargin=*]
  \item The first segment has length $1\leq m\leq L+1$.
  \item Any segment besides the first and last segment has length $\lfloor(1-c)L\rfloor\leq m\leq L$.\label{it:segment_length1}
  \item The last segment has length $1\leq m\leq L$, if $\R$ has more than one segment.\label{it:segment_length2}
\end{enumerate}
\end{lem}
\ifdefined\Compact{}
\else
\begin{prf}
The bound $m\geq1$ is trivial, since any segment is non-empty.
Let $[k+1,k+m]\in\R$ be an arbitrary segment, let $t_0$ be the value of variable \texttt t at the beginning of step $k+1$ and let $t$ be the value of variable \texttt t at the beginning of step $k+m$.
Since there is no rescale in steps $k+1, k+2, \dots, k+m-1$, we have $t=t_0+(m-1)d$.
We now substitute a lower (upper) bound on $t$ and an upper (lower) bound on $t_0$ into $m=\frac{t-t_0}d+1$ to obtain a lower (upper) bound on $m$.
\block{First segment. ---}
By Condition \ref{it:cond_s} we have $t_0\geq N$ and by Lemma \ref{lem:total_max} we have $t\leq T$.
Thus, $m\leq (T-N)/d+1=L+1$.
\block{Neither first nor last segment. ---}
A rescale took place in step $k$, thus at the end of step $k$ (at the beginning of step $k+1$) we have $t_0\geq N+d$ (see \FuncSty{Update}) and by Lemma \ref{lem:total_max} we have $t_0\leq d+N+cdL$.
Since there is a rescale in step $k+m$ we must have $t>T-d$ (see \FuncSty{Update}) and by Lemma \ref{lem:total_max} we obtain $t\leq T$.
Consequently, $m\leq (T-N-d)/d+1=L$ and $m>(T-d-(d+N+cdL))/d+1 = (1-c)L-1$, which implies $m\geq\lfloor(1-c)L\rfloor$.
\block{Last segment. ---}
When $\lvert\R\rvert>1$ the first and last segment are distinct.
The length of the last segment is maximal, when a rescale takes place in its last step.
Thus, $m\leq L$, just as in the previous case.
\end{prf}
\fi

\noindent
We can improve the upper bounds in Cases \ref{it:segment_length1} and \ref{it:segment_length2} of Lemma \ref{lem:segment_length} to $\lfloor(1-c)L\rfloor+2$, by a better lower bound on \texttt t (for $d\geq N$) after a rescale.
Unfortunately, this turns out to be of little use for our main result.
So we use the simple versions.
Finally, the lower bounds on the length of a segment from a rescale partition imply an upper bound on the number of such segments.

\begin{lem}\label{lem:segment_count}
Any realization of Algorithm~$\rfd$ with rescale partition $\R$ satisfies $\lvert \R\rvert-2\leq\frac{n}{\lfloor(1-c)L\rfloor}$.
\end{lem}
\ifdefined\Compact{}
\else
\begin{prf}
For $\lvert\R\rvert\leq2$ the bound trivially holds.
Now let $\lvert\R\rvert>2$.
By Lemma~\ref{lem:segment_length} all but the first and last segment have length at least $\lfloor(1-c)L\rfloor$.
Hence, $n>(\lvert\R\rvert-2)\lfloor(1-c)L\rfloor$ and dividing by $\lfloor(1-c)L\rfloor\geq1$ (by Condition \ref{it:cond_T}) concludes the proof.
\end{prf}
\fi

\section{Code Length Analysis}\label{sec:analysis}

\paragraph{(Partial) Redundancy Bounds.}
We are now ready to compare the performance of Algorithm $\rfd$ to a competing (ideal) model, namely an arbitrary, fixed probability distribution from $\P$.
The comparison criterion is the redundancy of Algorithm $\rfd$ compared to the competing model, i.\,e. the number of bits Algorithm $\rfd$ requires above the competitor to encode an arbitrary input sequence (assuming ideal encoding).
Later, we will enhance the competitor model to become a sequence of arbitrary probability distributions from $\P$.
We begin the analysis with a redundancy bound for a special realization, namely when $\lvert\R\rvert=1$.
Now consider a realization with only one segment in its rescale partition $\R$ of $[1,n]$.
There is at most one rescale in step $n$. 
However, this rescale doesn't affect $\rfd(x; x^k)$ for $0\leq k<n$.
In this simple scenario the probability assignment of Algorithm $\rfd$ collapses to
\begin{align}
  \rfd(x; x^k) = \frac{s_0(x) + d\cdot \lvert \{ i\mid x_i=x \text{ \upshape for } 1\leq i\leq k \} \rvert}{t_0+d \cdot k}
  \label{eq:rfd_simple}
  \text.
\end{align}
We now give an upper bound bound on the redundancy of Algorithm $\rfd$ in this setting.
To do so, we first state two lemmas.

\begin{lem}\label{lem:multinomial}
For non-negative integers $c(x)$, where $x\in\X$, with sum $n>0$ and any probability distribution $p\in\P$, we have $\log\binom{n}{c(1) , \dots , c(N)}\leq \sum_{x\in\X} c(x)\log \frac1{p(x)}$.
\end{lem}
\ifdefined\Compact{}
\else
\begin{prf}
W.\,l.\,o.\,g. we assume that $c(x)\geq1$ for $1\leq x\leq M$ and $c(x)=0$ for $M<x\leq N$, where $M\geq1$, since $n>0$.
Let $q(x) := c(x)/n$, the r.\,h.\,s. of the claimed inequality is
\begin{align}
  \sum_{x\in\X} c(x)\log \frac1{p(x)} = n (H(q) + D(q\parallel p)) \geq n H(q)
  \label{eq:multinomial1}
  \text.
\end{align}
By the Multinomial Theorem and by $\binom{n}{c(1),\dots,c(M)}=\binom{n}{c(1),\dots,c(N)}$ we get
\begin{align}
  1 = \left(\sum_{1\leq x\leq M} q(x)\right)^{\mathclap n} {\geq} \binom{n}{c(1),\dots, c(M)} \cdot\prod_{\mathclap{1\leq x\leq M}} q(x)^{c(x)} {=} \binom{n}{c(1),\dots,c(N)} 2^{-n H(q)}
  \label{eq:multinomial2}
 \text.
\end{align}
Combining \eqref{eq:multinomial1} and \eqref{eq:multinomial2} and rearranging completes the proof.
\end{prf}
\fi

\noindent
Notice that using Stirling's formula the above bound can be improved.
However, in the worst case the improvement is small, thus we omit it and use the simple version.
Now we proceed with an intermediate upper bound on the code length of Algorithm $\rfd$.

\begin{lem}\label{lem:rfd_multinomial_bound}
Consider a realization of Algorithm $\rfd$ with exactly one segment in its rescale partition.
Symbol $x$ occurs $c(x)$ times in $x^n$, $M$ denotes the number of distinct symbols in $x^n$ and let $s_0\leq s_0(x)$ for all $x\in\X$.
It holds that
\[
  \ell(x^n; \rfd)
    \leq \log {\textstyle \binom{n}{c(1) \,\dots\, c(N)}} + \left(M-1+\frac{t_0-s_0}d\right)\log(n)
    +\log\frac{t_0/d \cdot e^{t_0/d}}{(s_0 /d)^M M^{M-s_0/d}}
    \text.
\]
\end{lem}
\begin{prf}
We bound $p(x^n) := \prod_{1\leq k\leq n} \rfd(x_k; x^{k-1})$, given by \eqref{eq:rfd_simple}, from below by separately treating numerator and denominator and by combining afterwards.

\block{Numerator. ---}
Let $s(x; x^k) := s_0(x)/d + \lvert \{ i\mid x_i=x \text{ \upshape for } 1\leq i\leq k \} \rvert$ and $a:=s_0/d$.
W.\,l.\,o.\,g. we assume that symbols $1\leq x\leq M$ have non-zero frequency, i.\,e. for these we have $c(x)\geq1$.
This yields
\begin{align}
  {
  \prod_{1\leq k\leq n} s(x_k; x^{k-1})}
  &{= \prod_{x\in\X} \prod_{\substack{1\leq k\leq n,\\x_k=x}} s(x; x^{k-1}) 
  \geq a^M \prod_{x\in\X} \prod_{1\leq k<c(x)} (a + k)} \\
  &{= a^M \cdot \prod_{1\leq x\leq  M} \left( \frac{c(x)!}{c(x)} \prod_{1\leq k<c(x)} (1+a/k) \right)} \label{eq:prob_bound_num0}\\
  &{\geq \left(\frac{a M}{n}\right)^M \cdot \prod_{1\leq x\leq M} \left(c(x)!\right) \cdot \prod_{1\leq k\leq n-M} \frac{a+k}k}
  \text.
  \label{eq:prob_bound_num}
\end{align}
For the last step define $C_x := \sum_{y<x} (c(y)-1)$ and to \eqref{eq:prob_bound_num0} apply \mbox{$\prod_{1\leq x\leq M} c(x) \leq (n/M)^M$} (by the Arithmetic-Geometric-Mean inequality) and
\[
{
  \prod_{1\leq x\leq M} \prod_{1\leq l<c(x)} \left(1+\frac{a}l\right) \geq \prod_{1\leq x\leq M} \prod_{1\leq l<c(x)} \left(1+\frac{a}{l+C_x}\right) = \prod_{1\leq k\leq n-M} \frac{k+a}k}
  \text.
\]

\block{Denominator. ---}
We set $t(x^k) := b+k$, where $b := t_0/d$, and rearrange to obtain
\begin{align}
  {
  \prod_{1\leq k\leq n} t(x^{k-1})
  = b \cdot\prod_{1\leq k<n} (b + k)
  = \frac{b n!}n \cdot \prod_{1\leq k<n} \frac{b+k}k}
  \text.
  \label{eq:prob_bound_dnom}
\end{align}

\block{Combination. ---}
Clearly $\rfd(x; x^k) = s(x; x^k)/t(x^k)$.
We combine inequality \eqref{eq:prob_bound_num} and \eqref{eq:prob_bound_dnom}, note definition of $\binom{n}{c(1),\dots,c(M)}$ and obtain
\begin{align}
  {p(x^n)}
  &{\geq \frac{1}{\binom{n}{c(1),\dots,c(M)}} \cdot \frac{(a M)^M}{b} \cdot n^{-M+1} \cdot \prod_{1\leq k<n} \left( \frac{a+k}{b+k}\right) \cdot \prod_{n-M<k<n} \frac1{1+a/k}} \\
  &{\geq \frac{1}{\binom{n}{c(1),\dots,c(N)}} \cdot \frac{(a M)^M}{b} \cdot n^{-M+1} \cdot (e n)^{-(b-a)} \cdot (eM)^{-a}}
  \label{eq:rfd_multinomial_bound1}
\end{align}
The last step follows from $\binom{n}{c(1)\,\dots\,c(M)}=\binom{n}{c(1)\,\dots\,c(N)}$, from
\[
{
  \prod_{1\leq k< n} \frac{b+ k}{a+k}
  \leq \prod_{1\leq k\leq n} \left(1+\frac{b-a}{a+k}\right)\leq e^{(b-a) H_n} \leq e^{(b-a)\ln(e n)} = (e n)^{b-a}}
  \text,
\]
where $H_n := \sum_{1\leq k\leq n} k^{-1}$, and from
\[
{
  \prod_{n-M<k<n} \left(1+\frac a k\right) \leq \prod_{0<k\leq M}\left(1+\frac a k\right)\leq e^{a H_M} \leq e^{a \ln(e M)} = (eM)^a}
  \text.  
\]
Rearranging \eqref{eq:rfd_multinomial_bound1} and substituting the definitions of $a$ and $b$ yields the claim.
\end{prf}

\noindent
We proceed by combining the previous two lemmas.

\begin{prp}\label{prp:rfd_simple_redundancy_bound}
For any realization of Algoritm $\rfd$ with exactly one segment in its rescale partition and for any probability distribution $p\in\P$ it holds that
\[
  %\ell(x^n; \rfd)-\ell(x^n; p) \leq \frac{(N-1)d+t_0-1}d \log(n) + \log\frac{t_0 e^{t_0/d}}d + N\log(d)
  \ell(x^n; \rfd)-\ell(x^n; p) \leq \frac{(N-1)d+t_0-1}d \log(n) + \log\left(t_0/d\cdot e^{t_0/d}\right) + N\log(d)
  \text.
\]
\end{prp}
\begin{prf}
For the proof we first combine Lemma \ref{lem:multinomial} and Lemma \ref{lem:rfd_multinomial_bound}.
Finally we have $(s_0 /d)^M M^{M-s_0/d} \geq d^{-M} M^{M-1} \geq d^{-N}$, since $d\geq1$ and $s_0=1$ (the content of any array cell $\texttt s[x]$ is at least one, at any time) and $1\leq M\leq N$.
\end{prf}

\noindent
If we set $d=1$ and $t_0=N$ we resemble the Laplace estimator and get redundancy $2(N-1)\log n + O(1)$.
This is worse by a factor of $2$ compared to the optimal bound, however our bound holds for a strictly larger class of models.

\paragraph{First Main Result.}
Recall that in the previous paragraph, we assume at most one rescale in the last time step $n$.
We will now eliminate this limitation and proceed to the general case.
For this we need the following observation and the subsequent main technical lemma.

\begin{obs}\label{obs:split_rfd}
Consider subset $(k,k+m]$ of a segment from a rescale partition of an arbitrary realization of Algorithm~$\rfd$.
Conditions \ref{it:cond0}-\ref{it:cond1} hold for steps from $[1,n]$, thus these must hold for steps from $(k,k+m]$, as well.
Let us now define $y^m = x_{k+1}^{k+m}$ and assign label $s_0(x)$ to the content of array cell $\texttt s[x]$ and $t_0$ to content of variable \texttt t, both at the end of step $k$.
If we now execute Algorithm~$\rfd$ on input sequence $y^m$, with initial values $s_0(x)$ and $t_0$ (for function \FuncSty{Init}), we create a realization with rescale partition $\{[1,m]\}$.
In this situation we have $\rfd(x_{k+i}; x^{k-i-1})=\rfd(y_i; y^{i-1})$ for $1\leq i\leq m$, since Algorithm~$\rfd$ is deterministic.
Furthermore, there is at most one rescale in step $m$.
We conclude that for a subset $(k,k+m]$ of a segment from the rescale partition of an arbitrary realization of Algorithm~$\rfd$ we can construct an input $s_0(\cdot), t_0$ and $y^m$ for Algorithm~$\rfd$, s.\,t.
the resulting realization has rescale partition $\{[1,m]\}$ and $\ell(x_{k+1}^{k+m}; \rfd) = \ell(y^m; \rfd)$, where Proposition~\ref{prp:rfd_simple_redundancy_bound} holds for $\ell(y^m; \rfd)$.
\end{obs}

\begin{lem}\label{lem:rfd_segment_redundancy_bound}
Let $A:=\frac{N}d+cL$, let $p\in\P$ be an arbitrary probability distribution and for $z>0$ define function $r(z) := (N+A) \log(z) + \log((A+1) e^{A+1}) + N\log d$.
For a realization of Algorithm~$\rfd$ and rescale partition $\R_{\I}$ of $\I=[a,b]\subseteq[1,n]$ it holds that
\[
  \ell(x_{a}^{b}; \rfd)-\ell(x_a^b; p)\leq
    \lvert \R_{\I}\rvert r\left(\frac {\lvert {\I}\rvert}{\lvert\R_{\I}\rvert}\right) + (1-c)L\log(e m) + \log((1-c)L)
    \text,
\]
where $m$ is the length of the first segment of $\R_{\I}$.
\end{lem}
\begin{prf}
The plan for this proof is as follows. Clearly, we have
\begin{align}
  \ell(x_{a}^{b}; \rfd)-\ell(x_a^b; p) = \sum_{[j,k]\in\R_{\I}}\left( \ell(x_j^k; \rfd)-\ell(x_j^k; p)\right)
\end{align}
and by Observation~\ref{obs:split_rfd} we can bound every summand from the r.\,h.\,s. of the previous equation from above by Proposition~\ref{prp:rfd_simple_redundancy_bound}.
For this we distinguish between the first segment and subsequent segments from $\R_{\I}$ and sum the redundancy bounds afterwards.

\block{First segment. ---}
Let $s=[j,k]$ be the first segment of $\R_{\I}$.
By Observation~\ref{obs:split_rfd} and Proposition~\ref{prp:rfd_simple_redundancy_bound} with $\frac{t_0}d\leq \frac{T}d = \frac{N}d+L$ (due to Lemma~\ref{lem:total_max} and by \eqref{eq:L}), we get
\begin{align}
  \frac{\ell(x_j^k; \rfd)-\ell(x_j^k; p)}{\log e}
  &{<}
    \left(N+\frac N d+L\right)\ln(\lvert s\rvert) + \frac N d + L + \ln\left( \frac N d+L\right) %\\
     + N\ln(d)
\intertext{We bound $\ln\left( \frac N d+L\right)$ on the r.\,h.\,s. from above using $\frac N d+L\leq (1-c)L(\frac N d+cL+1)$, since $1\leq (1-c)L$ (by Condition~\ref{it:cond_T} and \eqref{eq:L}), and use $\frac N d + L = A+(1-c)L$:}
  \frac{\ell(x_j^k; \rfd)-\ell(x_j^k; p)}{\log e}
  &\leq
  \begin{multlined}[t]
    \left(N+A + (1-c)L\right)\ln(\lvert s\rvert) + A+(1-c)L  \\
    + \ln((1-c)L) +\ln( A+1)  + N\ln(d)
  \end{multlined}\\
  &<\frac{r(\lvert s\rvert)}{\log e} + (1-c)L\ln(e \lvert s\rvert)+ \ln((1-c)L)
  \text.
\end{align}

\block{Subsequent segment. ---}
Let $s=[j,k]$ be any but the first segment of $\R_{\I}$.
Analogous to the first segment, except $\frac{t_0}d\leq 1+\frac Nd+cL=A+1$ (due to Lemma~\ref{lem:total_max} and by \eqref{eq:L}), we get $\ell(x_j^k; \rfd)-\ell(x_j^k; p) \leq r(\lvert s\rvert)$.

\block{Summing. ---}
We get
$
  (1-c)L\log(e m)+ \log((1-c)L) + \lvert R_{\I}\rvert \sum_{s \in\R_{\I}} \frac1{\lvert \R_{\I}\rvert} r( \lvert s\rvert)
  %\text.
$
as an upper bound on the total redundancy.
Since $r(z)$ is concave, we may use Jensen's inequality on the rightmost sum of the bound and the claim immediately follows.
\end{prf}

\noindent
At this point the bulk of technical work is done.
As stated previously we now want to compare the code length of Algorithm~$\rfd$ to a sequence of probability distributions, that is:

\begin{dfn}
A piecewise stationary model \textsc{pws} for sequences of length $n$ is a tuple $(\S, (p_{\I})_{{\I}\in\S})$, where $\S$ is a partition of $[1,n]$ and $p_{\I}\in\P$ for all ${\I}\in\S$.
We define the shorthand $\textsc{pws}(x; x^{k}) := p_{\I}(x)$, where $k+1\in {\I}$ for segment ${\I}\in\S$.
\end{dfn}

\begin{lem}\label{lem:rescale_partition_sum}
For any realization of Algorithm~$\rfd$ with rescale partition $\R$ and any partition $\S$ of $[1,n]$ we have $\lvert\R\rvert\leq \sum_{\I\in\S} \lvert\R_{\I}\rvert\leq \lvert\R\rvert+\lvert\S\rvert-1$, where $\R_{\I}$ is the rescale partition of segment $\I\in\S$ induced by $\R$.
\end{lem}
\ifdefined\Compact{}
\else
\begin{prf}
Segment $\I\in\S$ has between $\lvert\R_{\I}\rvert-1$ and $\lvert\R_{\I}\rvert$ rescales.
Let $j$ be the last segment of $\S$ and distinguish:

\block{Case 1: There is a rescale in step $n$. ---}
We have $\lvert\R\rvert$ rescales in total and $\lvert\R_j\rvert$ rescales in segment $j$.
Hence, $\sum_{\I\in\S\setminus\{j\}}(\lvert\R_{\I}\rvert-1) + \lvert\R_j\rvert\leq \lvert\R\rvert$ and $\sum_{\I\in\S}\lvert\R_{\I}\rvert\geq\lvert\R\rvert$, the claim holds.

\block{Case 2: No rescale in step $n$. ---}
We have $\lvert\R\rvert-1$ rescales in total and $\lvert\R_j\rvert-1$ rescales in segment $j$,
i.\,e. $\sum_{\I\in\S}(\lvert\R_{\I}\rvert-1) \leq \lvert\R\rvert-1$ and $\lvert\R_j\rvert-1+\sum_{\I\in\S\setminus\{j\}}\lvert\R_{\I}\rvert\geq\lvert\R\rvert-1$, the claim holds.
\end{prf}
\fi

\noindent

Using Lemmas \ref{lem:rfd_segment_redundancy_bound} and \ref{lem:rescale_partition_sum} we now can get our first main result which relates redundancy of Algorithm~$\rfd$ to parameter $L$.
(Recall that $L=(T-N)/d$ essentially is an upper bound on the length of a segment from a rescale partition.)

\begin{thm}\label{thm:rfd_main1}
For any realization of Algorithm~$\rfd$ with rescale partition $\R$, any piecewise stationary model $\textsc{pws}$ for a sequence of length $n$ with partition $\S$ and function $r$, as defined in Lemma~\ref{lem:rfd_segment_redundancy_bound}, we have
\begin{multline}
  \ell(x^n; \rfd)-\ell(x^n; \textsc{pws})\leq
    \lvert\S\rvert\Big[ (1-c)L\log(e (L+1)) + \log((1-c)L) + r(L+1)\Big] \\
    + (\lvert \R\rvert-1) r(L+1) 
    \label{eq:rfd_main1}
    \text.
\end{multline}
\end{thm}
\begin{prf}
Let $\R_{\I}$ be the rescale partition of interval $\I\subseteq[1,n]$.
We obtain
\begin{multline}
\ell(x^n; \rfd)-\ell(x^n; \textsc{pws}) = \sum_{[a,b]\in\S} \left(\ell(x_a^b; \rfd) - \ell(x_a^b; \textsc{pws})\right)\\
  \leq \lvert S\rvert \Big[(1-c)L\log(e (L+1) ) + \log((1-c)L) \Big] + \sum_{\I\in\S} \lvert R_{\I}\rvert r\left(\frac {\lvert \I\rvert }{\lvert\R_{\I}\rvert}\right)
  \text,
\end{multline}
which follows from Lemma~\ref{lem:rfd_segment_redundancy_bound} and from $m\leq L+1$, by Lemma~\ref{lem:segment_length}.
To conclude the proof we define $R:=\sum_{\I\in\S} \lvert\R_{\I}\rvert$ and bound the rightmost sum from above,
\[
  R\sum_{\I\in\S} \frac{\lvert R_{\I}\rvert}R r\left(\frac {\lvert \I\rvert }{\lvert\R_{\I}\rvert}\right) \leq R\cdot r\left(\frac nR\right) \leq (\lvert\R\rvert+\lvert\S\rvert-1)\cdot r(L+1)
  \text.
\]
Step one follows from Jensen's inequality ($r$ is concave);
step two follows from $R\leq \lvert\R\rvert+\lvert\S\rvert-1$, by Lemma~\ref{lem:rescale_partition_sum}, from the fact that $r$ is increasing, from $n/R\leq n/{\lvert\R\rvert}$, again, by Lemma~\ref{lem:rescale_partition_sum}, and from $n/\lvert\R\rvert\leq L+1$. (The average length of a segment from $\lvert\R\rvert$ is at most the maximum segment length, which is $L+1$, by Lemma~\ref{lem:segment_length}.)
\end{prf}

\paragraph{Parameter Interaction.}
Let us discuss bound \eqref{eq:rfd_main1}.
By $r(L+1) = O\left( (N+cL) \log(L)\right)$ the r.\,h.\,s. of bound \eqref{eq:rfd_main1} is 
\begin{align}
  O\left(\lvert\S\rvert L\log(L) + \lvert\R\rvert (N+cL)\log(L) \right)
  \text.
  \label{eq:rfd_main1_O}
\end{align}
The redundancy essentially consists of two parts:
First, redundancy introduced by the complexity (i.\,e. number $\lvert\S\rvert$ of segments) of the competitor, secondly, redundancy due to rescaling.
Intuitively, it should be hard to go head to head with a sophisticated ($\lvert\S\rvert$ is large) competitor.
Bound \eqref{eq:rfd_main1} confirms this, every additional segment for the competitor costs $O(L\log L)$ bits.
Keep in mind that bound \eqref{eq:rfd_main1} holds for any competitor.
Hence, it holds for the competitor, which maximizes the redundancy, so we have a worst-case bound.
We now give a fatal scenario to interpret the redundancy of $O((N+cL)\log L)$ bits per rescale.
Consider a realization of Algorithm~$\rfd$, with one rescale.
(The example can be generalized to any number of rescales.)
Before the rescale Algorithm~$\rfd$ encountered the sequence $11\dots1$, and $22\dots2$ after the rescale.
Right after the rescale Algorithm~$\rfd$ will still assign high probability to letter $1$.
However, in this scenario it would be wise to discard the statistics from the previous segment.
The smaller parameter $c$, the better Algorithm~$\rfd$ will perform in this scenario, this attributes redundancy $O(cL\log L)$.
But when $c=0$, we still have to pay $O(N \log L)$ bits for adaption.
By combining this gives redundancy $O((N+cL) \log L)$ bits.
(Expressed more technically: By Proposition~\ref{prp:rfd_simple_redundancy_bound} the redundancy for the segment of length $m$ is $O((N+t_0)\log m)$, by Lemma~\ref{lem:segment_length} $m=O(L)$ and by Lemma~\ref{lem:total_max} $t_0=O(N+cL)$.)

As a tradeoff we set $c=\gamma/L$.
When $\gamma=O(1)$, this does not significantly worsen the (worst-case) performance \eqref{eq:rfd_main1_O} and keeping statistics from previous observations usually offers good performance in practice.
By $c=\gamma/L$ and by Lemma~\ref{lem:segment_count}, which implies $\lvert\R\rvert = O(\frac n{L-\gamma})$, we turn \eqref{eq:rfd_main1_O} into
\begin{align}
  O\left( \lvert\S\rvert L\log(L) + \frac{(N+\gamma)\log(L)}{L-\gamma} \cdot n \right)
  \label{eq:rfd_main1_O1}
  \text.
\end{align}
It is hard to tune parameter $L$, since this requires knowledge of desirable $\lvert\S\rvert$.
At this point we just give an example to illustrate the influence of $L$.
\begin{ex}\label{ex:L}
The choice of $L=\sqrt{n}$ and $c=\gamma/L$ for $\gamma=O(1)$ asymptotically minimizes \eqref{eq:rfd_main1_O1}, when $\lvert\S\rvert$ is $O(1)$.
We still guarantees sublinear redundancy, whenever $\lvert\S\rvert$ is $o({\sqrt n}/{\log n})$.
For the given parameter configuration we get redundancy $O(\lvert\S\rvert \sqrt n \log n)$.
\end{ex}

\paragraph{Second Main Result.}
Typically we do not know sequence length $n$ in advance, thus we cannot easily tune parameters $c$ and $L$ for sublinear redundancy for specific $\lvert\S\rvert$.
To iron out this deficit, we give a weaker version of Theorem~\ref{thm:rfd_main1}, with the following rationale:
It is unfair to compare Algorithm~$\rfd$ to a (sequence of) \emph{arbitrary} probability distributions, since Algorithm~$\rfd$ can assign at most probability $\frac{T-N+1}T<1$, but not probability $1$, to any letter.
Therefore the per-letter code length is bounded away from $0$.
So it is fairer, to choose a competitor based on probability distributions with probabilities bounded away from $1$.
This makes up our second main result.

\begin{thm}\label{thm:rfd_main2}
Let $\textsc{pws}=(\S, (p_{\I})_{\I\in\S})$ be a piecewise stationary model for a sequence of length $n$, s.\,t. for all $\I\in S$ probability distribution $p_{\I}$ has probability at most $1-\eps$ on any letter.
For any realization of Algorithm~$\rfd$, where $c=\gamma/L$ for integer $\gamma$ and $\delta := \frac{ r(L+1) }{\eps (L-\gamma)\log e }$, we have
\begin{multline}
  \ell(x^n; \rfd)-(1+\delta) \cdot\ell(x^n; \textsc{pws})
    \leq \lvert\S\rvert \cdot\Big[ (L-\gamma)\log(e (L+1)) +  \\[-6pt] + \log(L-\gamma) + r(L+1) \Big]+ r(L+1)
    \text.
    \label{eq:rfd_main2}
\end{multline}
\end{thm}
\begin{prf}
We combine Theorem~\ref{thm:rfd_main1} with $(1-c)L=L-\gamma$ and obtain
\begin{multline}
  \ell(x^n; \rfd)-\ell(x^n; \textsc{pws})\leq
    \lvert\S\rvert\cdot\Big[ (L-\gamma)\log(e (L+1)) + \log(L-\gamma) + r(L+1)\Big] + r(L+1) \\
    + (\lvert \R\rvert-2) \cdot r(L+1)
    \text.
    \label{eq:rfd_main2_1}
\end{multline}
It remains to show that $(\lvert\R\rvert-2)\cdot r(L+1)\leq \delta \cdot \ell(x^n; \textsc{pws})$.
To do so we combine Lemma~\ref{lem:segment_count} with $\lfloor(1-c)L\rfloor=L-\gamma$ and with $\ell(x^n; \textsc{pws})\geq -n \log (1-\eps) \geq \eps n \log e$ (since the maximum probability is bounded), this yields
\[
  (\lvert \R\rvert-2) \cdot r(L+1)
  \leq \frac{n \, r(L+1)}{L-\gamma}
  = \frac{ r(L+1)}{\eps (L-\gamma)\log e}\cdot \eps n \log e
  \leq \delta \cdot \ell(x^n; \textsc{pws})
  \text.
  \qedhere
\]
\end{prf}

\noindent
Theorem~\ref{thm:rfd_main2} guarantees that the code length of Algorithm~$\rfd$ will be within $(1+\delta)$ times the code length of the competing scheme plus $O(\lvert\S\rvert L\log L)$ bits of redundancy.
Ideally we want
\[
  \delta = \left[\left(N+\frac Nd+\gamma\right)\ln(L+1) + \frac Nd+\gamma+1 + \ln\left(\frac Nd+\gamma+1\right)\right]\frac1{\eps(L-\gamma)}
\]
to be as small as possible.
Indeed, for $1/\eps=o(L/\log L)$ (and constant $\gamma$ and $N$) $\delta$ decreases as $L$ increases.
So $\delta$ can be made arbitrarily small at the cost of increasing the constant on the r.\,h.\,s. of \eqref{eq:rfd_main2}.

\section{Conclusion}\label{sec:conclusion}
\vspace{-2pt}
In this work we revisited a well-known technique for probability estimation, which is based on relative symbol frequencies and a periodic frequency discount, Algorithm~$\rfd$.
We identified central parameters of Algorithm~$\rfd$ and carried out an analysis, based on realistic constraints on these parameters.
For the analysis we compared the difference in code length of Algorithm~$\rfd$ w.\,r.\,t. two competing schemes, in terms of an upper bound.
A sequence of arbitrary and constrained probability distributions served as competitors.
Results indicate that asymptotically Algorithm~$\rfd$ is close to either competitor.
(The proximity is a trade-off between the complexity of the competitor and the choice of parameters.)

The given bounds can be improved technically (e.\,g. by enhancing Lemma~\ref{lem:multinomial}).
An experimental study on Algorithm~$\rfd$ offers room for future research.
Furthermore, we believe that narrowing the parameter space ($d\geq N$, for instance) or narrowing the class of competitors (e.\,g. $\S=\R$) results in improved bounds.
It would be interesting to obtain redundancy bounds which depend on a more fine-grained characterization of PWS (here: only $\lvert\S\rvert$), such as the sum of a similarity measure for adjacent PWS partitions.
An important refinement of Algorithm~$\rfd$, is exponential aging of non-integer symbol frequencies, which is very successful in practice \cite{cm_ccp2011,ptw13,cts12}.
This approximately equals the situation where a rescale takes place in every step.
Another modification of Algorithm~$\rfd$ can be sketched as follows:
Perform the $k$-th rescale $k$ steps after the $(k-1)$-st rescale.
In this scheme there are $O(\sqrt n)$ rescales and we conjecture that the redundancy w.\,r.\,t. a piecewise stationary model with partition $\S$ is $O(\lvert\S\rvert\sqrt n\log n)$.
Our second main result opens up an interesting perspective on the design of finite state machine predictors:
When the parameters of Algorithm~$\rfd$ are fixed and independent of $n$, there is only a finite number of ``states'' $(\texttt s, \texttt t)$.
Thus, we can construct a finite state machine that simulates Algorithm~$\rfd$ and give theoretical guarantees on its performance.
(This addresses the third open problem given by Merhav and Feder in \cite{merhav98}.)
\vspace{-2pt}
\paragraph{Acknowledgement.}
The author thanks Martin Dietzfelbinger, Martin Aumueller and the anonymous reviewers for helpful comments and suggestions.

% \bibliographystyle{plain}
% \bibliography{references}{}

\bibliographystyle{plain}
\bibliography{references}{}

\end{document}